\documentclass{article}

\usepackage{epsfig}
\usepackage{subfigure}
\usepackage{calc}
\usepackage{graphicx}

\usepackage{amsfonts}
\usepackage{amsmath}
\usepackage{amssymb}
\usepackage{latexsym}

\newtheorem{theorem}{Theorem}[section]
\newtheorem{lemma}[theorem]{Lemma}

\newtheorem{remark}[theorem]{Remark}

\newtheorem{proposition}[theorem]{Proposition}

\begin{document}

\title{Modelling Nonlinear Sequence Generators in terms of Linear Cellular Automata}
\date{}
\author{Amparo F\'{u}ster-Sabater$^{(1)}$ and Dolores de la Gu\'{\i}a-Mart\'{\i}nez$^{(2)}$\\
{\small (1) Instituto de F\'{\i}sica Aplicada, C.S.I.C.}\\
{\small Serrano 144, 28006 Madrid, Spain} \\
{\small amparo@iec.csic.es}\\
{\small (2) Centro T\'{e}cnico de Inform\'{a}tica, C.S.I.C.} \\
{\small Pinar 19, 28006 Madrid, Spain} \\
{\small lola@cti.csic.es }}

\maketitle

\begin{abstract}
In this work, a wide family of LFSR-based sequence generators, the
so-called Clock-Controlled Shrinking Generators (CCSGs), has been
analyzed and identified with a subset of linear Cellular Automata
(CA). In fact, a pair of linear models describing the behavior of
the CCSGs can be derived. The algorithm that converts a given CCSG
into a CA-based linear model is very simple and can be applied to
CCSGs in a range of practical interest. The linearity of these
cellular models can be advantageously used in two different ways:
(a) for the analysis and/or cryptanalysis of the CCSGs and (b) for
the reconstruction of the output sequence obtained from this kind
of generators.


\end{abstract}

\section{Introduction}
\footnotetext{Research supported by Ministerio de Educaci\'{o}n y
Ciencia
(Spain) under grant SEG2004-02418 and SEG2004-04352-C04-03. \\
Applied Mathematical Modelling. Volume 31, Issue 2, pp. 226-235. February 2007. \\
DOI:10.1016/j.apm.2005.08.013 } \noindent Cellular Automata (CA)
are discrete dynamic systems characterized by a simple structure
but a complex behavior, see \cite{Da}, \cite{Mar}, \cite{Nan},
\cite{Wolfram1} and \cite{Wolfram3}. They are built up by
individual elements, called cells, related among them in many
varied ways. CA have been used in application areas so different
as physical system simulation, biological process, species
evolution, socio-economical models or test pattern generation.
Their simple, modular, and cascable structure makes them very
attractive for VLSI implementations. CA can be characterized by
several parameters which determine their behavior e.g. the number
of states per cell, the function $\Phi$ (the so-called
\textit{rule}) under which the cellular automaton evolves to the
next state, the number of neighbor cells which are included in
$\Phi$, the number of preceding states included in $\Phi$, the
geometric structure and dimension of the automaton (the cells can
be arranged on a line or in a square or cubic lattice in two,
three or more dimensions), ... etc.

On the other hand, Linear Feedback Shift Registers (LFSRs)
\cite{Go} are electronic devices currently used in the generation
of pseudorandom sequences. The inherent simplicity of LFSRs, their
ease of implementation, and the good statistical properties of
their output sequences turn them into natural building blocks for
the design of pseudorandom sequence generators with applications
in spread-spectrum communications, circuit testing,
error-correcting codes, numerical simulations or cryptography.

CA and LFSRs are special forms of a more general mathematical
structure: finite state machines \cite{Stone}. In recent years,
one-dimensional CA have been proposed as an alternative to LFSRs
(\cite{Bao}, \cite{Blackburn}, \cite{Nan} and \cite{Wolfram2}) in
the sense that every sequence generated by a LFSR can be obtained
from one-dimensional CA too. Pseudorandom sequence generators
currently involve several LFSRs combined by means of nonlinear
functions or irregular clocking techniques (see \cite{Menezes},
\cite{Rueppel}). Then, the question that arises in a natural way
is: are there one-dimensional CA able to produce the sequence
obtained from any LFSR-based generator? The answer is yes and, in
fact, this paper considers the problem of given a particular
LFSR-based generator how to find one-dimensional CA that reproduce
its output sequence. More precisely, in this work it is shown that
a wide class of LFSR-based nonlinear generators, the so-called
Clock-Controlled Shrinking Generators (CCSGs) \cite{Kanso}, can be
described in terms of one-dimensional CA configurations. The
automata here presented unify in a simple structure the above
mentioned class of sequence generators. Moreover, CCSGs that is
generators conceived and designed as \textit{nonlinear} models are
converted into \textit{linear} one-dimensional CA. Once the
generators have been linearized, all the theoretical background on
linear CA found in the literature can be applied to their analysis
and/or cryptanalysis. The conversion procedure is very simple and
can be realized in a range of practical interest.

The paper is organized as follows: in section 2, basic concepts
e.g. one-dimensional CA, CCSGs or the Cattel and Muzio cellular
synthesis method are introduced. A simple algorithm to determine
the pair of CA corresponding to a particular shrinking generator
and its generalization to Clock-Controlled Shrinking Generators
are given in sections 3 and 4, respectively. A simple approach to
the reconstruction of the generated sequence that exploits the
linearity of the CA-based model is presented in section 5.
Finally, conclusions in section 6 end the paper.

\section{Basic Structures}
In the following subsections, we introduce the general
characteristics of the basic structures we are dealing with:
one-dimensional cellular automata, the shrinking generator and the
class of clock-controlled shrinking generators. Throughout the
work, only binary CA and LFSRs will be considered. In addition,
all the LFSRs we are dealing with are maximal-length LFSRs whose
output sequences are \textit{PN-}sequences \cite{Go}.

\subsection{One-Dimensional Cellular Automata}
One-dimensional cellular automata can be described as
\textit{n}-cell registers \cite{Da}, whose cell contents are
updated at the same time according to a particular rule; that is
to say a \textit{k}-variable function denoted by $\Phi$. If the
function $\Phi$ is a linear function, so is the cellular
automaton. When $k$ input binary variables are considered, then
there is a total of $2^{k}$ different neighbor configurations.
Therefore, for cellular automata with binary contents there can be
up to $2^{2^{k}}$ different mappings to the next state. Moreover,
if $k=2r+1$, then the next state $x_{i}^{t+1}$ of the cell
$x_{i}^{t}$ depends on the current state of $k$ neighbor cells
$x_{i}^{t+1}=\Phi (x_{i-r}^{t},\ldots ,x_{i}^{t},\ldots
,x_{i+r}^{t}) \;\; (i=1, ..., n)$.

CA are called \textit{uniform} whether all cells evolve under the
same rule while CA are called \textit{hybrid} whether different
cells evolve under different rules. At the ends of the array, two
different boundary conditions are possible: \textit{null automata}
when cells with permanent null contents are supposed adjacent to
the extreme cells or \textit{periodic automata} when extreme cells
are supposed adjacent.

In this paper, all the automata considered will be one-dimensional
null hybrid CA with $k=3$ and linear rules 90 and 150. Such rules
are described as follows:

\begin{center}
Rule 90  \qquad \qquad \qquad \qquad \qquad \qquad Rule 150\\
$x_{i}^{t+1}=x_{i-1}^{t}\oplus x_{i+1}^{t}$ \qquad \qquad \qquad
$x_{i}^{t+1}=x_{i-1}^{t}\oplus x_{i}^{t}\oplus x_{i+1}^{t}$
\end{center}

\setlength{\tabcolsep}{4pt}
\begin{table}
\begin{center}
\caption{An one-dimensional null hybrid linear cellular automaton
of $10$ cells with rule 90 and rule 150 starting at a given
initial state} \label{table:headings1}
\begin{tabular}{cccccccccc}
\hline\noalign{\smallskip}
$\;\;90\;$ & $\;\;150$ & $\;\;150$ & $\;\;150$ & $\;\;90\;$ & $\;\;90\;$ & $\;\;150$ & $\;\;150$ & $\;\;150$ & $\;\;90\;$ \\
\noalign{\smallskip} \hline \noalign{\smallskip}
$\;0$ & $\;0$ & 0 & 1 & 1 & 1 & 0 & 1 & 1 & 0\\
$\;0$ & $\;0$ & 1 & 0 & 0 & 1 & 0 & 0 & 0 & 1 \\
$\;0$ & $\;1$ & 1 & 1 & 1 & 0 & 1 & 0 & 1 & 0 \\
$\;1$ & $\;0$ & 1 & 1 & 1 & 0 & 1 & 0 & 1 & 1 \\
$\;0$ & $\;0$ & 0 & 1 & 1 & 0 & 1 & 0 & 0 & 1 \\
$\;0$ & $\;0$ & 1 & 0 & 1 & 0 & 1 & 1 & 1 & 0 \\
$\;0$ & $\;1$ & 1 & 0 & 0 & 0 & 0 & 1 & 0 & 1 \\
$\;1$ & $\;0$ & 0 & 1 & 0 & 0 & 1 & 1 & 0 & 0 \\
$\;0$ & $\;1$ & 1 & 1 & 1 & 1 & 0 & 0 & 1 & 0 \\
$\;1$ & $\;0$ & 1 & 1 & 0 & 1 & 1 & 1 & 1 & 1 \\
$\;\vdots$ & $\;\vdots$ & \vdots & \vdots & \vdots & \vdots & \vdots & \vdots & \vdots & \vdots \\
\\
\hline
\end{tabular}
\end{center}
\end{table}
For an one-dimensional null hybrid cellular automaton of length
$n=10$ cells, configuration rules
$(\,90,150,150,150,90,90,150,150,150,90\,)$ and initial state
$(0,0,0,1,1,$ $1,0,1,1,0)$, Table \ref{table:headings1}
illustrates the formation of its output sequences (binary
sequences read vertically) and the succession of states (binary
configurations of 10 bits read horizontally). For the above
mentioned rules, the different states of the automaton are grouped
in closed cycles. The number of different output sequences for a
particular cycle is $\leq n$ as the same sequence (although
shifted) may appear simultaneously in different cells. At the same
time, all the sequences in a cycle will have the same period and
linear complexity \cite{Mar} as well as any output sequence of the
automaton can be produced at any cell provided that we get the
right state cycle.

\subsection{The Shrinking Generator}
The shrinking generator is a binary sequence generator \cite{Co}
composed by two LFSRs : a control register, called $R_{1}$, that
decimates the sequence produced by the other register, called
$R_{2}$. We denote by $L_j\; (j=1,2)$ their corresponding lengths
and by $P_j(x)\in GF(2)[x]\; (j=1,2)$ their corresponding
characteristic polynomials \cite{Go}.

The sequence produced by the LFSR $R_{1}$, that is $\{a_{i}\}$,
controls the bits of the sequence produced by $R_{2}$, that is
$\{b_{i}\}$, which are included in the output sequence $\{c_{j}\}$
(\textit{the shrunken sequence}), according to the following rule
$P$:

\begin{enumerate}
\item  If $a_{i}=1\Longrightarrow c_{j}=b_{i}$

\item  If $a_{i}=0\Longrightarrow b_{i}$ is discarded.
\end{enumerate}

A simple example illustrates the behavior of this structure.

\textit{Example 1: }Let us consider the following LFSRs:

\begin{enumerate}
\item  Shift register $R_{1}$ of length $L_{1}=3$, characteristic
polynomial $P_{1}(x)=1+x^{2}+x^{3}$ and initial state
$IS_{1}=(1,0,0)$. The sequence generated by $R_{1}$ is
$\{a_{i}\}=\{1,0,0,1,1,1,0\}$ with period $T_1=2^{L_1}-1=7$.

\item  Shift register $R_{2}$ of length $L_{2}=4$, characteristic
polynomial $P_{2}(x)=1+x+x^{4}$ and initial state
$IS_{2}=(1,0,0,0)$. The sequence generated by $R_{2}$ is
$\{b_{i}\}=\{1,0,0,0,1,0,0,1,1,0,1,0,1,1,1\}$ with period
$T_2=2^{L_2}-1=15$.
\end{enumerate}

The output sequence $\{c_{j}\}$ is given by:
\begin{itemize}
\item  $\{a_{i}\}$ $\rightarrow $ $1\;0\;0\;1\;1\;1\;0\;1\;0\;0\;1\;1\;1\;0%
\;1\;0\;0\;1\;1\;1\;0\;1\;.....$

\item  $\{b_{i}\}$ $\rightarrow $ $\hspace{0.02cm}1\;\underline{0}\;\underline{0}\;0\;1\;0\;%
\underline{0}\;1\;\underline{1}\;\underline{0}\;1\;0\;1\;\underline{1}\;1\;\underline{1}%
        \;\underline{0}\;0\;0\;1\;\underline{0}\;0\;.....$

\item  $\{c_{j}\}$ $\rightarrow $
$1\;0\;1\;0\;1\;1\;0\;1\;1\;0\;0\;1\;0\;.....$
\end{itemize}

The underlined bits \underline{0} or \underline{1} in $\{b_{i}\}$
are discarded. In brief, the sequence produced by the shrinking
generator is an irregular decimation of $\{b_{i}\}$ from the bits
of $\{a_{i}\}$.

According to \cite{Co}, the period of the shrunken sequence is
\begin{equation}
T=(2^{L_{2}}-1)2^{(L_{1}-1)}
\end{equation}
and its linear complexity \cite{Rueppel}, notated $LC$, satisfies
the following inequality
\begin{equation}
L_{2} \thinspace 2^{(L_{1}-2)}<LC\leq L_{2} \thinspace
2^{(L_{1}-1)}.
\end{equation}
In addition, it can be proved \cite{Co} that the output sequence
has some nice distributional statistics too. Therefore, this
scheme is suitable for practical implementation of stream cipher
cryptosystems and pattern generators.

\subsection{The Clock-Controlled Shrinking Generators}
The Clock-Controlled Shrinking Generators constitute a wide class
of clock-controlled sequence generators \cite{Kanso} with
applications in cryptography, error correcting codes and digital
signature. An CCSG is a sequence generator composed of two LFSRs
notated $R_{1}$ and $R_{2}$. The parameters of both registers are
defined as those of subsection 2.2. At any time $t$,the control
register $R_{1}$ is clocked normally while the second register
$R_{2}$ is clocked a number of times given by an integer
decimation function notated $X_t$. In fact, if $A_0(t),\,
A_1(t),\,\ldots,\, A_{L_{1}-1}(t)$ are the binary cell contents of
$R_{1}$ at time $t$, then $X_t$ is defined as
\begin{equation}
X_{t}=1+2^0 A_{i_0}(t) + 2^1 A_{i_1}(t) + \ldots + 2^{w-1}
A_{i_{w-1}}(t)
\end{equation}
where $i_0,\,i_1,\,\ldots,\, i_{w-1}\in \{0,\, 1,\, \ldots,
\,L_{1}-1\}$ and $0 < w \leq L_{1}-1$.

In this way, the output sequence of an CCSG is obtained from a
double decimation. First, $\{b_{i}\}$ the output sequence of
$R_{2}$ is decimated by means of $X_t$ giving rise to the sequence
$\{b'_{i}\}$. Then, the same decimation rule $P$, defined in
subsection 2.2, is applied to the sequence $\{b'_{i}\}$. Remark
that if $X_t \equiv 1$ (no cells are selected in $R_1$), then the
proposed generator is just the shrinking generator. Let us see a
simple example of CCSG.

\textit{Example 2: } For the same LFSRs defined in the previous
example and the function $X_{t}=1+2^0 A_{0}(t)$ with $w=1$, the
decimated sequence $\{b'_{i}\}$ is given by:

\begin{itemize}
\item  $\{b_{i}\}$ $\rightarrow $ $1\;\underline{0}\;0\;0\;1\;\underline{0}\;0\;\underline{1}\;1\;\underline{0}\;1\;0\;\underline{1}\;1\;1%
\;1\;\underline{0}\;0\;\underline{0}\;1\;\underline{0}\;0\;1\;\underline{1}\;0\;1\;0\;\underline{1}\;1\;\underline{1}\;1\;.....$

\item  $\;X_{t}\;$ $\rightarrow $
$2\;1\;1\;2\;2\;2\;1\;2\;1\;1\;2\;2\;2\;1\;2\;1\;1\;2\;2\;.....$

\item  $\{b'_{i}\}$ $\rightarrow $ $1\;0\;0\;1\;0\;1\;%
1\;0\;1\;1\;1\;0\;1\;0\;1\;0%
        \;1\;0\;1\;1\;.....$
\end{itemize}

According to the decimation function $X_t$, the underlined bits
\underline{0} or \underline{1} in $\{b_{i}\}$ are discarded in
order to produce the sequence $\{b'_{i}\}$. Then the output
sequence $\{c_{j}\}$ of the CCSG output sequence is given by:

\begin{itemize}
\item  $\{a_{i}\}$ $\rightarrow $ $1\;0\;0\;1\;1\;1\;0\;1\;0\;0\;1\;1\;1\;0%
\;1\;0\;0\;1\;1\;1\;0\;1\;.....$

\item  $\{b'_{i}\}$ $\rightarrow $ $\hspace{0.02cm}1\;\underline{0}\;\underline{0}\;1\;0\;1\;%
\underline{1}\;0\;\underline{1}\;\underline{1}\;1\;0\;1\;\underline{0}\;1\;\underline{0}%
        \;\underline{1}\;0\;1\;1\;.....$

\item  $\{c_{j}\}$ $\rightarrow $
$1\;1\;0\;1\;0\;1\;0\;1\;1\;0\;1\;1\;.....$
\end{itemize}

The underlined bits \underline{0} or \underline{1} in $\{b'_{i}\}$
are discarded.

In brief, the sequence produced by an CCSG is an irregular double
decimation of the sequence generated by $R_2$ from the function
$X_t$ and the bits of $R_1$. This construction allows one to
generate a large family of different sequences by using the same
LFSR initial states and characteristic polynomials but modifying
the decimation function. Period, linear complexity and statistical
properties of the generated sequences by CCSGs have been
established in \cite{Kanso}.

\subsection{Cattel and Muzio Synthesis Algorithm}
The Cattell and Muzio synthesis algorithm \cite{Ca} presents a
method of obtaining two CA (based on rules 90 and 150)
corresponding to a given polynomial. Such an algorithm takes as
input an irreducible polynomial $Q(x) \in GF(2)[x]$ defined over a
finite field and computes two reversal linear CA whose output
sequences have $Q(x)$ as characteristic polynomial. Such CA are
written as binary strings with the following codification: $0$ =
rule $90$ and $1$ = rule $150$. The theoretical foundations of the
algorithm can be found in \cite{Ca1}. The total number of
operations required for this algorithm is listed in
\cite{Ca}(Table II, page 334). It is shown that the number of
operations grows linearly with the degree of the polynomial, so
the method does not suffer from any sort of exponential blow-up.
The method is efficient for all practical applications (e.g. in
1996 finding a pair of length $300$ CA took 16 CPU seconds on a
SPARC 10 workstation). For cryptographic applications, the degree
of the irreducible (primitive) polynomial is $ L_{2}\approx 64$,
so that the consuming time is negligible.

Finally, a list of One-Dimensional Linear Hybrid Cellular Automata
of Degree Through 500 can be found in \cite{Ca2}.

\section{CA-Based Linear Models for the Shrinking Generator}

In this section, an algorithm to determine the pair of
one-dimensional linear CA corresponding to a given shrinking
generator is presented. Such an algorithm is based on the
following results:

\begin{lemma}
The characteristic polynomial of the shrunken sequence is of the
form $P(x)^{N}$, where $P(x)\in GF(2)[x]$ is a $L_{2}$-degree
polynomial and $N$ is an integer satisfying the inequality
$2^{(L_{1}-2)}<N\leq 2^{(L_{1}-1)}$.
\end{lemma}
\textit{Sketch of proof.}$\;$ The idea of the proof consists in
demonstrating the uniqueness of the polynomial $P(x)$ that defines
the linear recurrence relation satisfied by $\{c_{j}\}$ for both
the upper and lower bounds on the linear complexity. The values of
such bounds are given in equation (2).
\hfill $\Box$

\begin{lemma}
Let $P_{2}(x)\in GF(2)[x]$ be the characteristic polynomial of
$R_2$ and let $\alpha$ be a root of $P_{2}(x)$ in the extension
field $GF(2^{L_{2}})$. Then, $P(x)\in GF(2)[x]$ is the
characteristic polynomial of \textit{cyclotomic coset}
$2^{L_{1}}-1$, that is

\begin{equation}
P(x) =(x+\alpha^{E})(x+\alpha^{2E})\ldots
(x+\alpha^{2^{L_{1}-1}E})
\end{equation}
being $E$ an integer given by
\begin{equation}
E =2^{0}+2^{1}+\ldots +2^{L_{1}-1} \;.
\end{equation}
\end{lemma}
\textit{Sketch of proof.}$\;$ The shrunken sequence can be written
as an interleaved sequence \cite{Gong} made out of an unique
\textit{PN}-sequence repeated $2^{(L_{1}-1)}$ times where
$2^{(L_{1}-1)}$ is the number of $1's$ in a full period of
$\{a_{i}\}$. Such a \textit{PN}-sequence is obtained from
$\{b_{i}\}$ taking digits separated a distance $2^{L_1}-1$. That
is the \textit{PN}-sequence is the characteristic sequence
associated with the cyclotomic coset $2^{L_1}-1$ whose
characteristic polynomial is $P(x)$.\hfill $\Box$

Remark that $P(x)$ depends exclusively on the characteristic
polynomial of the register $R_{2}$ and on the length $L_{1}$ of
the register $R_{1}$. In addition, the polynomial $P(x)$ will be
the input to the Cattell and Muzio synthesis algorithm \cite{Ca}.
Based on such an algorithm, the following result is derived:
\begin{proposition}
Let $Q(x)\in GF(2)[x]$ be a polynomial defined over a finite field
and let $s_1$ and $s_2$ two binary strings codifying the two
linear CA obtained from the Cattell and Muzio algorithm. Then, the
two binary strings corresponding to the polynomial $Q(x)\cdot
Q(x)$ are:
 \[
  S'_{i} = {S_i}*{S_i^{*}}\;\;\; i=1, 2
  \]
where $S_i$ is the binary string $s_i$ whose least significant bit
has been complemented, $S_i^{*}$ is the mirror image of $S_i$ and
the symbol $*$ denotes concatenation.
\end{proposition}
\textit{Sketch of proof.}$\;$ The result is just a generalization
of the Cattell and Muzio synthesis algorithm, see \cite{Ca} and
\cite{Ca1}. The concatenation is due to the fact that rule $90$
($150$) at the end of the array in null automata is equivalent to
two consecutive rules $150$ ($90$) with identical sequences.\hfill
$\Box$

According to the previous results, the following linearization
algorithm is introduced:

\bigskip
\textbf{Input:} A shrinking generator characterized by two LFSRs,
$R_{1}$ and $R_{2}$, with their corresponding lengths, $L_{1}$ and
$L_{2}$, and the characteristic polynomial $P_{2}(x)$ of the
register $R_{2}$.

\begin{description}
\item  \textit{Step 1:} From $L_{1}$ and $P_{2}(x)$, compute the
polynomial $P(x)$ as
\begin{equation}
P(x) =(x+\alpha^{E})(x+\alpha^{2E})\ldots
(x+\alpha^{2^{L_{1}-1}E})
\end{equation}
with $E =2^{0}+2^{1}+\ldots +2^{L_{1}-1}$.

\item  \textit{Step 2:} From $P(x)$, apply the Cattell and Muzio
synthesis algorithm \cite{Ca} to determine the two linear CA (with
rules 90 and 150), notated $s_i$, whose characteristic polynomial
is $P(x)$.

\item  \textit{Step 3:} For each $s_i$ separately, we proceed:
\begin{description}
  \item \textit{3.1} Complement its least significant bit. The resulting binary string is notated $S_i$.

  \item \textit{3.2} Compute the mirror image of $S_i$, notated $S_i^{*}$, and
  concatenate both strings
  \[
  S'_{i} = S_i*S_i^{*}\;.
  \]
  \item \textit{3.3} Apply steps $3.1$ and $3.2$ to each $S'_{i}$ recursively $L_{1}-1$ times.
\end{description}

\end{description}

\textbf{Output: } Two binary strings of length $n=L_2
\;2^{L_{1}-1}$ codifying the linear CA corresponding to the given
shrinking generator.

\begin{remark}
The characteristic polynomial of the register $R_{1}$ is not
needed. Thus all the shrinking generators with the same $R_2$ but
different registers $R_1$ (all of them with the same length $L_1$)
can be modelled by the same pair of one-dimensional linear CA.
\end{remark}
\begin{remark}
It can be noticed that the computational requirements of the
linearization algorithm  are minimum. In fact, it just consists in
the application of the Cattell and Muzio synthesis algorithm whose
consuming time is negligible plus $(L_1-1)$ concatenations of
binary strings. Both procedures can be carried out on a simple PC.
\end{remark}

In any case, thanks to this simple algorithm a linear model
producing the output sequence of the shrinking generator is
obtained. In order to clarify the previous steps a simple
numerical example is presented.

\bigskip
\textbf{Input:} A shrinking generator characterized by two LFSRs
$R_{1}$ of length $L_{1}=3$ and $R_{2}$ of length $L_{2}=5$ and
characteristic polynomial $P_{2}(x)=1+x+x^{2}+x^{4}+x^{5}$. Now
$E=2^3 - 1$

\begin{description}
\item  \textit{Step 1:} $P(x)$ is the characteristic polynomial of
the cyclotomic \textit{coset 7}. Thus,
\[
P(x) =1+x^{2}+x^{5}\;.
\]

\item  \textit{Step 2:} From $P(x)$ and applying the Cattell and
Muzio synthesis algorithm, two reversal linear CA whose
characteristic polynomial is $P(x)$ can be determined. Such CA are
written in binary format as:
\begin{center}
$
\begin{array}{ccccc}
0 & 1 & 1 & 1 & 1 \\
1 & 1 & 1 & 1 & 0
\end{array}
$
\end{center}

\item  \textit{Step 3:} Computation of the required pair of CA.

For the first automaton:
\begin{center}
$
\begin{array}{cccccccccccccccccccc}
0 & 1 & 1 & 1 & 1 \\
0 & 1 & 1 & 1 & 0 & 0 & 1 & 1 & 1 & 0\\
0 & 1 & 1 & 1 & 0 & 0 & 1 & 1 & 1 & 1 & 1 & 1 & 1 & 1 & 0 & 0 & 1
& 1 & 1 & 0\;\; \mbox{(final automaton)}
\end{array}
$
\end{center}
For the second automaton:
\begin{center}
$
\begin{array}{cccccccccccccccccccc}
1 & 1 & 1 & 1 & 0 \\
1 & 1 & 1 & 1 & 1 & 1 & 1 & 1 & 1 & 1\\
1 & 1 & 1 & 1 & 1 & 1 & 1 & 1 & 1 & 0 & 0 & 1 & 1 & 1 & 1 & 1 & 1
& 1 & 1 & 1\;\; \mbox{(final automaton)}
\end{array}
$
\end{center}
For each automaton, the procedure in \textit{Step 3} has been
carried out twice as $L_{1}-1 = 2$.

\bigskip
\textbf{Output: } Two binary strings of length $n = 20$ codifying
the required CA.

\end{description}

In this way, we have obtained a pair of linear CA among whose
output sequences we can obtain the shrunken sequence corresponding
to the given shrinking generator. Remark that the model based on
CA is a linear one. In addition, for each one of the previous
automata there are state cycles where the shrunken sequence is
generated at any one of the cells.

\section{CA-Based Linear Models for the Clock-Controlled Shrinking Generators}

In this section, an algorithm to determine the pair of
one-dimensional linear CA corresponding to a given CCSG is
presented. Such an algorithm is based on the following results:

\begin{lemma}
 The characteristic polynomial of the output sequence of a CCSG
is of the form $P'(x)^{N}$, where $P'(x)\in GF(2)[x]$ is a
$L_{2}$-degree polynomial and $N$ is an integer satisfying the
inequality $2^{(L_{1}-2)}<N\leq 2^{(L_{1}-1)}$.
\end{lemma}
\textit{Sketch of proof.}$\;$ The idea of the proof is analogous
to that one developed in Lemma 3.1. \hfill $\Box$

Remark that, according to the structure of the CCSGs, the
polynomial $P'(x)$ depends on the characteristic polynomial of the
register $R_{2}$, the length $L_{1}$ of the register $R_{1}$ and
the decimation function $X_t$. Before, $P(x)$ was the
characteristic polynomial of the \textit{cyclotomic coset} $E$,
where $E=2^{0}+2^{1}+\ldots +2^{L_{1}-1}$ was a fixed separation
distance between the digits drawn from the sequence $\{b_{i}\}$.
Now, this distance $D$ is variable and is a function of $X_t$. The
computation of $D$ gives rise to the following result:

\begin{lemma}
Let $P_{2}(x)\in GF(2)[x]$ be the characteristic polynomial of
$R_2$ and let $\alpha$ be a root of $P_{2}(x)$ in the extension
field $GF(2^{L_{2}})$. Then, $P'(x)\in GF(2)[x]$ is the
characteristic polynomial of \textit{cyclotomic coset} $D$, where
$D$ is given by
\begin{equation}
D=2^{L_1-w}\; (\sum \limits_{i=1}^{2^w}i) \;
-1=(1+2^w)\;2^{L_1-1}\;-1.
\end{equation}
\end{lemma}

\textit{Sketch of proof.}$\;$ The idea of the proof is analogous
to that one developed in Lemma 3.2. In fact, the distance $D$ can
be computed taking into account that the function $X_t$ takes
values in the interval $[1,\;2,\;\ldots,\;2^w]$ and the number of
times that each one of these values appears in a period of the
output sequence is given by $2^{L_1-w}$. A simple computation,
based on the sum of the terms of an arithmetic progression,
completes the sketch.\hfill $\Box$

From the previous results, it can be noticed that the algorithm to
determine the CA corresponding to a given CCSG is analogous to
that one developed in section 3; just the expression of $E$ in
equation (4) must be here replaced by the expression of $D$ in
equation (7). A simple numerical example is presented.

\bigskip
\textbf{Input:} A CCSG characterized by: Two LFSRs $R_{1}$ of
length $L_{1}=3$ and $R_{2}$ of length $L_{2}=5$ and
characteristic polynomial $P_{2}(x)=1+x+x^{2}+x^{4}+x^{5}$ plus
the decimation function $X_{t}=1+2^0 A_{0}(t) + 2^1 A_{1}(t) + 2^2
A_{2}(t)$ with $w=3$.

\begin{description}
\item  \textit{Step 1:} $P'(x)$ is the characteristic polynomial
of the cyclotomic \textit{coset D}. Now $D \equiv 4\;mod\;31$,
that is we are dealing with the cyclotomic coset $1$. Thus, the
corresponding characteristic polynomial is:
\[
P'(x) =1+x+x^{2}+x^{4}+x^{5}\;.
\]

\item  \textit{Step 2:} From $P'(x)$ and applying the Cattell and
Muzio synthesis algorithm, two reversal linear CA whose
characteristic polynomial is $P'(x)$ can be determined. Such CA
are written in binary format as:
\begin{center}
$
\begin{array}{ccccc}
1 & 0 & 0 & 0 & 0 \\
0 & 0 & 0 & 0 & 1
\end{array}
$
\end{center}

\item  \textit{Step 3:} Computation of the required pair of CA.

For the first automaton:
\begin{center}
$
\begin{array}{cccccccccccccccccccc}
1 & 0 & 0 & 0 & 0 \\
1 & 0 & 0 & 0 & 1 & 1 & 0 & 0 & 0 & 1\\
1 & 0 & 0 & 0 & 1 & 1 & 0 & 0 & 0 & 0 & 0 & 0 & 0 & 0 & 1 & 1 & 0
& 0 & 0 & 1\;\; \mbox{(final automaton)}
\end{array}
$
\end{center}
For the second automaton:
\begin{center}
$
\begin{array}{cccccccccccccccccccc}
0 & 0 & 0 & 0 & 1 \\
0 & 0 & 0 & 0 & 0 & 0 & 0 & 0 & 0 & 0\\
0 & 0 & 0 & 0 & 0 & 0 & 0 & 0 & 0 & 1 & 1 & 0 & 0 & 0 & 0 & 0 & 0
& 0 & 0 & 0\;\; \mbox{(final automaton)}
\end{array}
$
\end{center}
For each automaton, the procedure in \textit{Step 3} has been
carried out twice as $L_{1}-1 = 2$.

\bigskip
\textbf{Output: } Two binary strings of length $n = 20$ codifying
the required CA.
\end{description}
\begin{remark}
From a point of view of the CA-based linear models, the shrinking
generator or any one of the CCGS are entirely analogous. Thus, the
fact of introduce an additional decimation function does neither
increase the complexity of the generator nor improve its
resistance against cryptanalytic attacks since both kinds of
generators can be linearized by the same class of CA-based models.
\end{remark}

\section{A Simple Approach to the Output Sequence Reconstruction for this Class of Sequence Generators}

Since CA-based linear models describing the behavior of CCSGs have
been derived, a cryptanalytic attack that exploits the weaknesses
of these models has been also developed. It consists in
reconstructing the CCSG output sequence from an amount of such a
sequence (the intercepted subsequence). The key idea of this
attack is based on the study of the repeated sequences in the
automata under consideration and the relative shifts among such
sequences. In fact, the sequence at a extreme cell of the
automaton is repeated on average once out of $L_2$ cells. In order
to determine these shifts, the algorithm of Bardell \cite{Bar} to
phase-shift analysis of CA is applied. The approach is composed by
several steps:

\begin{itemize}
\item  Step 1: The portion of $M$ intercepted bits of the output
sequence is placed at the most right (left) cell of one of the
automata. This provides shifted portions of the same output
sequence produced at different cells. The lengths of these
subsequences are (on average) $ (M-L_2), (M-2L_2), (M-3L_2),
\ldots , (M-pL_2)$ where $p=\lfloor M/L_2\rfloor$.

\item  Step 2: The locations of the different cells that generate
the same output sequence as well as the relative shifts among
these sequences are detected via Bardell's algorithm.

\item  Step 3: Repeat Steps 1 and Step 2 for every one of the
subsequences obtained above.
\end{itemize}

Summing up the contributions of the bits provided by each
automaton, we obtain that the total number of bits reconstructed
is
\begin{equation}\label{equation:11}
N_T \approx M p^2=M (M/L_2)^2
\end{equation}

We know not only this number of bits but also the precise location
of such bits along the sequence. Notice that we have two different
CA plus an additional pair of CA corresponding to the reverse
version of the output sequence (the pair associated to the
reciprocal polynomial of $P_2(x)$). In addition, for each
automaton the intercepted M-bit sequence can be placed either at
the most right cell or the most left cell producing different
locations of the same sequence. Thus, each one of the different
automata will contribute to the reconstruction of the output
sequence with a number of bits given by the equation
(\ref{equation:11}). Moreover, remark that the output sequence for
these generators is an interleaved sequence \cite{Gong} made out
of a fixed \textit{PN-}sequence. Hence, the portions of the
reconstructed subsequence allow us to fix the starting point of
many of these \textit{PN-}sequences. The rest of the bits of each
\textit{PN-}sequence can be easily derived.

Once the previous steps are accomplished, the original output
sequence can be reconstructed by concatenating all different
reconstructed subsequences.

Finally, let us see a simple example of application of Bardell's
algorithm.

\textit{Example 3: } Let us consider a cellular automaton with the
following characteristics:

\begin{itemize}
\item Number of cells $n=10$

\item Automaton under study in binary format: $0011001100$

\item Characteristic polynomial $(1+x+x^{3}+x^{4}+x^{5})^{2}\;.$

\end{itemize}

  Let $S$ be the shift operator defined on
  $X_{i}\;(i=1,\ldots ,10)
$, the state of the \textit{i}-th cell , such as follows:
\begin{eqnarray*}
SX_{i}(t)= X_{i}(t+1)\;.
\end{eqnarray*}
Thus, the corresponding difference equation system for the
previous automaton can be written as follows:

\[
SX_{1} =X_{2} \;\;\;\;\;\; SX_{2} =X_{1}+X_{3} \;\;\;\; \ldots
\;\;\;\; SX_{10} =X_{9}\;.
\]

Next expressing each $X_{i}$ as a function of $X_{10}$, we obtain
the following system:
\begin{eqnarray*}
X_{1} &=& (S^9+S^4+S^3+S^2+S+1)X_{10} \\
X_{2} &=& (S^8+S^6+S^5+S^4+S^3+S+1) X_{10} \\
\vdots  \\
X_{9} &=& (S)X_{10}\;.
\end{eqnarray*}

Analogous results can be obtained expressing each $X_{i}$ as a
function of $X_{1}$. Now taking logarithms in both sides of the
equalities,
\begin{eqnarray*}
log(X_{1})&=&log(S^9+S^4+S^3+S^2+S+1)+log(X_{10}) \\
log (X_{2})&=&log(S^8+S^6+S^5+S^4+S^3+S+1)+ log(X_{10}) \\
\vdots \\
 log
(X_{9})&=&log(S)+log(X_{10})\;.
\end{eqnarray*}

The base of the logarithm is $R(S)$ and the values of the
logarithms are integers over a finite domain. According to the
Bardell's algorithm, we determine the integers $m$ (if there
exist) such that $S^{m}\thinspace mod \thinspace R(S)$ equal the
different polynomials in $S$ included in the above system. For
instance,
\begin{eqnarray*}
S^{26}\thinspace mod \thinspace R(S) = S^2+1\;.
\end{eqnarray*}
Or simply, $S^{26}=S^{2}+1$ and $26\log (S)=\log (S^{2}+1)$ with
$log (S) \equiv 1$. Now substituting in the previous system, the
following equations can be derived:
%\begin{multicols}{2}
\begin{eqnarray*}
log (X_{9}) - log (X_{10}) &=& 1 \\
log (X_{8}) - log (X_{10}) &=& 26 \\
log (X_{4}) - log (X_{10}) &=& 6
\end{eqnarray*}

\begin{eqnarray*}
log (X_{2}) - log (X_{1}) &=& 1 \\
log (X_{3}) - log (X_{1}) &=& 26 \\
log (X_{7}) - log (X_{1}) &=& 6\;.
\end{eqnarray*}
%\end{multicols}

The phase-shifts of the outputs 9, 8 and 4 relative to cell 10 are
1, 26 and 6 respectively. Similar values are obtained in the other
group of cells, that is cells 2, 3 and 7 relative to cell 1. The
other cells generate different sequences. Further contributions to
phase-shift analysis of CA based on 90/150 rules can be found in
\cite{Nan1} and \cite{Cho}.
\section{Conclusions}

A wide family of LFSR-based sequence generators, the so-called
Clock-Controlled Shrinking Generators, has been analyzed and
identified with a subset of linear cellular automata. In this way,
sequence generators conceived and designed as complex nonlinear
models can be written in terms of simple linear models. An easy
algorithm to compute the pair of one-dimensional linear hybrid
cellular automata that generate the CCSG output sequences has been
derived. A cryptanalytic approach based on the phase-shift of
cellular automata output sequences is proposed. From the obtained
results, we can create linear cellular automata-based models to
analyse/cryptanalyse the class of clock-controlled generators.

\end{document}